\documentclass[a4paper,12pt]{article}

\usepackage{bbold}
\usepackage{float}
\usepackage{cancel}
\usepackage{subfig}
\usepackage{amsmath, amssymb, setspace,cite}
\usepackage{graphicx}
\usepackage{hyperref}
\usepackage{color}
\usepackage{soul}
\usepackage{xcolor}
\usepackage{multirow}

\vspace{0.6cm}
\date{\today} 

\usepackage{tabularx}
\usepackage{longtable} 
\usepackage{multicol}
\usepackage{multirow}

\renewcommand{\arraystretch}{1.2}

\begin{document}

\hfill {\tt CERN-TH-2024-150}

\def\thefootnote{\fnsymbol{footnote}}

\begin{center}
\Large\bf\boldmath
\vspace*{1.cm} Theoretical implications for a new measurement of $K_L\to \pi^0 \ell\bar\ell$
\end{center}
\vspace{0.6cm}

\begin{center}
G.~D'Ambrosio$^{1}$\footnote{Electronic address: gdambros@na.infn.it}, 
A.M. ~Iyer$^{2}$\footnote{Electronic address: iyerabhishek@physics.iitd.ac.in}, F.~Mahmoudi$^{3,4,5}$\footnote{Electronic address: nazila@cern.ch}, 
S. Neshatpour$^{3}$\footnote{Electronic address: s.neshatpour@ip2i.in2p3.fr}\\
\vspace{0.6cm}
{\sl $^1$INFN-Sezione di Napoli, Complesso Universitario di Monte S. Angelo,\\ Via Cintia Edificio 6, 80126 Napoli, Italy}\\[0.4cm]
{\sl $^2$Department of Physics, Indian Institute of Technology Delhi,\\ Hauz Khas, New Delhi-110016, India}\\[0.4cm]
{\sl $^3$Universit\'e Claude Bernard Lyon 1, CNRS/IN2P3, \\
Institut de Physique des 2 Infinis de Lyon, UMR 5822, F-69622, Villeurbanne, France}\\[0.4cm]
{\sl $^4$Theoretical Physics Department, CERN, CH-1211 Geneva 23, Switzerland}\\[0.4cm]
{\sl $^5$Institut Universitaire de France (IUF), 75005 Paris, France }\\[0.4cm]
\end{center}
\renewcommand{\thefootnote}{\arabic{footnote}}
\setcounter{footnote}{0}

\vspace{1.cm}
\begin{abstract}
Kaon physics is at an important experimental juncture with respect to the ongoing measurements of several observables. This work will build on the existing status by formulating different phenomenological analyses corresponding to different paths that may lie ahead. 
Beginning with the golden channels, $K^+\rightarrow\pi^+\nu\bar\nu$ and $K_L\rightarrow\pi^0\nu\bar\nu$, the paper will eventually cast the spotlight on the importance of a precise measurement of \mbox{BR($K_L\rightarrow\pi^0\ell\bar\ell$)}. 
The phenomenological analyses involve sequentially adding kaon physics observables at the projected final precision of their respective measurements to the global fit. 
 More specifically, we consider three different scenarios with different sets of observables assumed at their final precisions.
Beginning with \mbox{BR($K^+\rightarrow\pi^+\nu\bar\nu$)} and BR($K_L\rightarrow\pi^0\nu\bar\nu$), we sequentially add BR($K_L\rightarrow\pi^0 e\bar e$) and BR($K_L\rightarrow\pi^0 \mu\bar\mu$) to the global fit. The evolution of the result from one scenario to the next makes a strong case for the consideration of future measurement of BR($K_L\rightarrow\pi^0\ell\bar\ell$).

\end{abstract}

\section{Introduction}
Kaon physics experiments have reached a critical crossroads that deserves investigation of possible avenues.  There are specific observables in the kaon sector that have been under intense experimental scrutiny to further augment our understanding of the Standard Model (SM). They include the golden channels: BR($K^+\rightarrow\pi^+\nu\bar\nu$) at NA62 \cite{NA62:2021zjw} and BR($K_L\rightarrow\pi^0\nu\bar\nu$) at KOTO \cite{Ahn:2018mvc,KOTO:2020prk}, the rare decay modes BR($K^+\rightarrow\pi^+\mu\bar\mu$) at NA62 \cite{NA62:2022qes} and $K_S\rightarrow\mu\bar \mu$ at LHCb~\cite{LHCb:2020ycd}. These analyses were preceded by several past endeavours that are inclusive of those at E787~\cite{E787:2002qfb,E787:2004ovg} and E949~\cite{E949:2004uaj,E949:2007xyy,E949:2008btt,BNL-E949:2009dza} for $K^+\to\pi^+\nu\bar\nu$, at E865~\cite{E865:1999ker} and NA48/2 \cite{NA482:2009pfe} for BR($K^+\rightarrow\pi^+ e\bar e$) and at NA48/2\cite{NA482:2010zrc} for BR($K^+\rightarrow\pi^+\mu\bar \mu$).  Each of them left a specific imprint on our understanding of the extent of New Physics (NP) beyond the Standard Model. However, it has become increasingly relevant to conduct a phenomenological study that would not only present a unified picture but also paint the different phenomenological goals that could be achieved based on the future evolution of these experiments.

In the first study of its kind for kaon decays, a global analysis was undertaken using a $(V-A)\otimes(V-A)$ type operator basis \cite{DAmbrosio:2022kvb}. The connections between the kaon decays with neutrinos and other charged leptons were facilitated by the $SU(2)$ symmetry that links the corresponding Wilson coefficients ($\delta C_9=-\delta C_{10}$) for the decay modes. By means of reasonable assumptions for the projected sensitivity for the measurement of different observables for the experiments, it illustrated the importance of pursuing a long-term kaon physics program.

While the study in \cite{DAmbrosio:2022kvb} was comprehensive in its own right, it also had the potential to expand its frontiers.  The consideration of operators beyond the $(V-A)\otimes(V-A)$ type would also open the doors for new observables that could be proposed for measurements. In particular, they include right-handed, scalar and tensor operators.
A novel prescription on obtaining bounds on the scalar operators from the $K^+\rightarrow\pi^+\ell\bar\ell$ was proposed in \cite{DAmbrosio:2024rxv}. A comprehensive global analysis using different combinations of operators is currently in progress.
Another important ingredient is the consideration of the possible impact on a precise measurement for $K_L\rightarrow\pi^0\ell\bar\ell$ on the global fits. This mode offers one of the best opportunities to test direct CP-violation. While the SM prediction of the branching ratio is of the order of $\sim 10^{-11}$, the current experimental bounds are roughly one order of magnitude away \cite{KTeV:2003sls,KTEV:2000ngj}. This is characteristic of both of the lepton modes $(\ell = e,\mu)$. Given the fact that experimental bounds and computed SM values are relatively close, it offers a strong motivation to consider its measurement in an upcoming experiment.  
 
This constitutes the primary goal of the paper where it will plant the seed for a serious deliberation of the measurement of BR($K_L\rightarrow\pi^0\ell\bar\ell$) at the proposed upgrade of KOTO \textit{viz.} KOTO\mbox{-}II \cite{Nanjo:2023xvj,NanjoTalkKaon2022}. Borrowing the methodology in \cite{DAmbrosio:2022kvb}, the paper will illustrate the extent to which the parameter space can be reduced giving projections guided by anticipated precision levels at the conclusion of NA62 and KOTO\mbox{-}II experiments. While the global fit will include all observables as those in \cite{DAmbrosio:2022kvb}, the projections will include the proposed improved precision for only a subset of them. They include BR($K^+\rightarrow\pi^+\nu\bar\nu$) with 15$\%$ projected precision at NA62, BR($K_L\rightarrow\pi^0\nu\bar\nu$) with 25$\%$ projected precision at KOTO\mbox{-}II and BR($K_L\rightarrow\pi^0\ell\bar\ell$) also with 25$\%$ projected precision at KOTO\mbox{-}II. In order to understand the impact of each of these precisions on the eventual global fit, the numerical analysis is divided into three scenarios where observables are progressively added. \textit{Scenario~1} only includes improved future measurements of BR($K^+\rightarrow
\pi^+\nu\bar\nu$) and BR($K_L\rightarrow\pi^0\nu\bar\nu$). Thereafter, \textit{Scenario 2} and \textit{Scenario 3} sequentially add BR($K_L\rightarrow\pi^0 e\bar e$) and BR($K_L\rightarrow\pi^0 \mu\bar\mu$), respectively, at their expected precisions to the global fits. The progressive reduction in the available parameter space of the Wilson coefficients in going from the present experimental status to \textit{Scenarios 1-3} makes a strong case for the measurement of BR($K_L\rightarrow\pi^0\ell\bar\ell$) with the specified precision. In all scenarios, we have maintained the theoretical precision at the current level (see e.g., Refs.~\cite{Anzivino:2023bhp,Neshatpour:2024lij} for the status of potential theoretical advancements).

The paper is organised as follows: In Section \ref{sec:sec2} we define the effective operator formalism and the corresponding observables that are considered for the analyses. The numerical strategy is outlined in Section \ref{sec:sec3} and the impact of the existing status of different kaon observables on the parameter space of the Wilson coefficients is illustrated. In Section \ref{sec:sec4}, we provide a detailed insight into the results obtained for all three scenarios. We conclude in Section \ref{sec:sec5} by identifying the future theoretical and new experimental prospects that mandate attention.

\section{Theoretical Methodology}
\label{sec:sec2}
The global analyses  follow the methodology established in \cite{DAmbrosio:2022kvb}, where the  effective Hamiltonian parameterising the $s\rightarrow d$ transitions is given as:

\begin{equation}\label{eq:Heff}
\mathcal{H}_{\rm eff}=-\frac{4G_F}{\sqrt{2}}\lambda_t^{sd}\frac{\alpha_e}{4\pi}\sum_k C_k^{\ell}O_k^{\ell}\,,
\end{equation}
where $\lambda_t^{sd}\equiv V^*_{ts}V_{td}$. In general, the summation over the index \textit{$k$} represents the inclusion of all types of operators. In this analysis, however, we restrict ourselves to the following three types of operators:
\begin{align}\nonumber
&{O}_9^{\ell} = (\bar{s} \gamma_\mu P_L d)\,(\bar{\ell}\gamma^\mu \ell)\,,
&&{O}_{10}^{\ell} = (\bar{s} \gamma_\mu P_L d)\,(\bar{\ell}\gamma^\mu\gamma_5 \ell)\,,&& {O}_L^{\ell} = (\bar{s} \gamma_\mu P_L d)\,(\bar{\nu}_\ell\,\gamma^\mu(1-\gamma_5)\, \nu_\ell)\,,
\label{eq:operators}
\end{align}
with $P_L=(1-\gamma_5)/2$ and the Wilson coefficients include both the SM and the NP contribution: $C^\ell_k=C^{\ell,SM}_k+\delta C^{\ell}_k$. The presence of a $SU(2)_L$ gauge symmetry permits the correlation of the Wilson coefficients between the charged and the neutral leptons.
This assumption also leads to the following relation between the Wilson coefficients in the chiral basis: $\delta C^\ell_9=-\delta C^\ell_{10}=\delta C^\ell_L$. Thus, all of the results will only be presented in terms of the respective $\delta C^\ell_L$ for the leptons.
This particular choice of operators facilitates a direct comparison with the results of \cite{DAmbrosio:2022kvb} thereby highlighting the potential impact of new observables with the prospect of KOTO\mbox{-}II~\cite{JParcKaon2024}.  

These operators can influence a large set of $s\rightarrow d$ observables, a subset of which is considered for the numerical analyses and listed below: 
\begin{itemize}
    \item \textbf{$\boldsymbol{K^+\rightarrow\pi^+ \nu\bar\nu}$ and $\boldsymbol{K_L\rightarrow\pi^0 \nu\bar\nu}$}: The presence of neutrinos in the final state of the decay of the kaon makes the long-distance contributions subdominant. In the SM, the mode is driven almost entirely by short-distance physics and this is a potential window to contributions beyond the SM. 
    The computation of the branching fraction involves a sum over the three neutrino species. In the language of Wilson coefficients, there is a possibility of both constructive/destructive interference between the  $\delta C^{e,\mu\,\tau}_L$. 
    As a result, even a measurement consistent with the SM does not necessarily imply the absence of NP contributions as potential lepton flavour universality violating (LFUV) effects could cancel each other in such a manner that a SM value is measured. This can readily be seen in Fig.~\ref{fig:GoldenChannels_reducedErrs} (in Appendix A) where a range of NP contributions results in the prediction of BR($K^+ \to \pi^+\nu\bar\nu$) in agreement with the SM (the donut shape). 
    The ongoing phase of the measurement of the respective branching fraction of these modes \cite{NA62:2021zjw,Ahn:2018mvc} makes it interesting from a phenomenological perspective\footnote{Very recently, after the completion of our study, the NA62 collaboration presented for the first time a measurement of $\text{BR}(K^+ \to \pi^+ \nu\bar\nu)$ with a statistical significance of $5\sigma$~\cite{Swallow2024}. This has not lead to any significant impact on our conclusions regarding future prospects.}.  The NA62 has the sensitivity to probe the SM value of BR($K^+\rightarrow\pi^+ \nu\bar\nu$) with a $15\%$ final precision~\cite{Anzivino:2023bhp, NA62:2856997}. While the ongoing analyses at KOTO is not expected to reach the SM value for the BR($K_L\rightarrow\pi^0 \nu\bar\nu$), the proposed KOTO\mbox{-}II can probe it with a $25\%$ sensitivity \cite{JParcKaon2024}.

    \item \textbf{$\boldsymbol{K^+\rightarrow\pi^+ \ell\bar\ell}$}: In the SM, the dominant contribution to this decay mode is through the single photon exchange due to $K\rightarrow \pi\gamma^*$ and is driven by the vector form factor ($W(z)=a_++b_+z$) \cite{DAmbrosio:1998gur,DAmbrosio:2018ytt,DAmbrosio:2019xph}.
    This is a multifaceted decay mode as its consideration either individually ($\ell=e$ or $\mu$) or together can open different doors for New Physics explorations. In the case of the former, the study of the decay spectrum and measurement of forward-backward asymmetry can be a portal to non-SM-type operators \cite{DAmbrosio:2024rxv}. In the case of the latter, comparing the structure of $W(z)$ for the electron and muon can provide concrete hints for deviation in the lepton flavour universality hypothesis by means of the computation of the difference $a^e_+-a^\mu_+$ \cite{Crivellin:2016vjc,DAmbrosio:2022kvb}. The electron mode was measured at the E865 \cite{E865:1999ker} and NA48/2 \cite{NA482:2009pfe} experiments. For the measurements in the muon mode, there is both an existing dataset from NA48/2 \cite{NA482:2010zrc}, and one ongoing in NA62 \cite{NA62:2022qes}. This mode will be considered in the context of LFUV and its existing precision will be kept in the fits for all projected scenarios.

    \item \textbf{$\boldsymbol{K_{L,S}\rightarrow\mu\bar\mu}$}: A common feature to both of the decay modes is the dominant long-distance contribution.   The amplitude for $K_L\rightarrow \mu\bar \mu$ is driven by the CP-odd contributions. This facilitates an interference between the long-distance (LD) and the short-distance (SD) contributions leading to a non-negligible value of its branching fraction in the SM. Consequently,  
    the decay $K_L\rightarrow\mu\bar\mu$ is characterised by a very precise measurement of its branching fraction \cite{PDG2020}. However, there is yet an unresolved ambiguity on the SM contribution owing to an undetermined sign in the long-distance contribution. 
     At the current level of theoretical precision, the sign of the long-distance contribution does not have a significant impact on the global fits.
    For this paper, we choose the positive sign (LD$+$) for the LD contribution \cite{DAmbrosio:2022kvb, DAmbrosio:2023irq} (corresponding to destructive interference between short- and long-distance contributions).
    The sign of the LD contribution is discussed in  \cite{DAmbrosio:1996kjn,DAmbrosio:1997eof,Isidori:2003ts,Gerard:2005yk,Hoferichter:2023wiy,DAmbrosio:2017klp}. 

    On the other hand, the amplitude for $K_S\rightarrow\mu\bar\mu$ is driven by the CP-even contributions \cite{Chobanova:2017rkj,Dery:2021vql,Dery:2021mct}. Thus, any SD physics of the form (SM-like) considered here will be suppressed and thus leading to a very small branching fraction. 
    Any enhancement in the branching fraction is not possible for reasonable values of the Wilson coefficients of the operators considered here. At present, this mode is characterised by an upper bound which is roughly two orders of magnitude away from its SM value\cite{LHCb:2020ycd}.  With  300 fb$^{-1}$ of data \cite{Chobanova:2020vmx}, the LHCb is expected to reach an upper bound $6.4\times 10^{-12}$ at 95$\%$ CL on the branching fraction \cite{LHCb:2018roe,LHCb:2020ycd,Diego}. Thus, this mode will have the least influence on the projected fits. 

    \item \textbf{$\boldsymbol{K_L\rightarrow\pi^0 \ell\bar\ell}$}: These modes are characterised by four different contribution: 1) A direct CP-violating contribution due to short-distance physics \cite{Buchalla:2003sj,Gilman:1979ud,Ecker:1987hd,Flynn:1988gy}, 2) an indirect CP-violating contribution induced due to $K^0-\bar K^0$ oscillation\cite{DAmbrosio:1998gur}, 3) an interference term resulting from the two mentioned contributions \cite{DAmbrosio:1998gur,Isidori:2004rb,Mescia:2006jd,Mescia:2007kn}, and 4) CP-conserving piece due to $\gamma\gamma\rightarrow\ell\bar\ell$ rescattering \cite{Ecker:1987fm,Cappiello:1988yg,Cappiello:1992kk,Cohen:1993ta,Morozumi:1988vy}. This makes this mode a promising testing ground for direct CP-violation \cite{DAmbrosio:1998gur}. The existing status of their measurement is roughly one order of magnitude away from their respective SM predictions\cite{KTeV:2003sls,KTEV:2000ngj}. KOTO\mbox{-}II will be expected to play a crucial role in reaching the SM expectation. We consider a potential measurement of both these modes at 25$\%$ precision. 
\end{itemize}

Table \ref{tab:data} provides a numerical summary of this section with the values for the precisions used in this analysis. 
A theoretical exposition for each of the decay modes is given in \cite{DAmbrosio:2022kvb} where the computation of the SM and the corresponding theoretical uncertainties are detailed. That  analysis serves as a benchmark based on which the proposal in this paper is formulated and detailed in the next section.

\section{Numerical Methodology}
\label{sec:sec3}
For our study, we consider a global fit to all rare kaon decays mentioned in the previous section using a frequentist approach:
\begin{align}
\chi^2(\delta C_L^\ell) = \sum_{i,j} \left(O_i^{\rm th}(\delta C_L^\ell) - O_i^{\rm exp}\right) C^{-1}_{i,j} \left(O_j^{\rm th}(\delta C_L^\ell) - O_j^{\rm exp}\right)
\end{align}
as implemented in the publicly available SuperIso program~\cite{Mahmoudi:2007vz,Mahmoudi:2008tp,Mahmoudi:2009zz,Neshatpour:2021nbn,Neshatpour:2022fak}. Here \( C_{i,j} \) is the covariance matrix which entails the experimental and theoretical uncertainties and correlations, and \( O_i^{\rm th}(\delta C_L^\ell) \) and \( O_i^{\rm exp} \) are the theoretical and experimental values for the $i$-th observable, respectively. As mentioned in the previous section and indicated by \( O_i^{\rm th}(\delta C_L^\ell) \), we only consider NP effects in \(\delta C_L^\ell = \delta C_9 = -\delta C_{10}\) in this work. However, this can be further expanded to also include contributions from scalar and pseudoscalar operators (e.g., see~\cite{Mescia:2006jd,DAmbrosio:2024rxv} for the impact of the latter contributions in $K\to \pi \ell\ell$ decays).
We adopt the numerical prescription of \cite{DAmbrosio:2022kvb} to serve two purposes: First, to highlight the role of the individual observables to the global fit. Secondly, it will also emphasise the importance of future experiments in narrowing down the parameter space of Wilson coefficients of the operators under consideration. This analysis would thus present the state of the art for the future precision of these Wilson coefficients. For simplicity, we adopt the following relation between the semileptonic Wilson coefficients
\begin{equation*}
    \delta C^e_L\neq \delta C^\mu_L =\delta C^\tau_L \,.
\end{equation*}
This reduces the dimensionality of the space from 3 to 2.
The current status and the expected sensitivity are given in Table \ref{tab:data}. In order to facilitate comparison with the analyses of \cite{DAmbrosio:2022kvb}, the same CKM inputs have been used in the computation of the SM predictions in Table \ref{tab:data}.

\begin{table}[!t]
\vspace{-5mm}
\renewcommand{\arraystretch}{1.39}
\begin{center}
\setlength\extrarowheight{1pt}
\scalebox{0.65}{
\hspace*{-2mm}
\begin{tabular}{|llll|c
|}\hline\hline
\bf{Observable} & \bf{SM prediction}& \bf{Experimental result} & \bf{Reference}&   \textbf{Precision for projections} \\ \hline
%%%
BR$(K^+\to \pi^+\nu\bar\nu)$    & $(7.86 \pm 0.61)\times 10^{-11}$  & $(10.6^{+4.0}_{-3.5} \pm 0.9 ) \times 10^{-11}$ & ~\cite{NA62:2021zjw}   & 15\% \cite{JParcKaon2024} \\
%%%
BR$(K^0_L\to \pi^0\nu\bar\nu)$  & $(2.68 \pm 0.30) \times 10^{-11}$ & $ <1.99\times 10^{-9}$ @$90\%$ CL & ~\cite{KOTO_ICEHP_2024} & $25\%$ \cite{JParcKaon2024}\\
%%%
LFUV($a_+^{\mu\mu}-a_+^{ee}$)&\multicolumn{1}{c}{0}&$-0.014\pm 0.016$&~\cite{DAmbrosio:2018ytt,NA62:2022qes} & Current  \\
BR$(K_L\to \mu\bar\mu)$ ($+$)   & $(6.82^{+0.77}_{-0.29})\times 10^{-9}$    & \multirow{2}{*}{$(6.84\pm0.11)\times 10^{-9}$} & \multirow{2}{*}{\cite{ParticleDataGroup:2024cfk}} &  \multirow{2}{*}{Current } \\
%%%
BR$(K_L\to \mu\bar\mu)$ ($-$)   &  $ (8.04^{+1.47}_{-0.98})\times 10^{-9}$      &  &&
\\
%%%
\multirow{2}{*}{BR$(K_S\to \mu\bar\mu)$}         & \multirow{2}{*}{$(5.15\pm1.50)\times 10^{-12}$}    & $ < 2.1(2.4)\times 10^{-10}$ @$90(95)\%$ CL & \multirow{2}{*}{~\cite{LHCb:2020ycd}}  & \multirow{2}{*}{$<6.4\times10^{-12}$ @$95\%$ CL (LHCb@300 fb$^{-1}$~\cite{LHCb:2018roe,Diego})} \\
& & ~~~~$\left( 0.9^{+0.7}_{-0.6}\times 10^{-10} \right)$ &  & \\ 
%%%
BR$(K_L\to \pi^0 e\bar e)(+)$         & $(3.46^{+0.92}_{-0.80})\times 10^{-11}$    & \multirow{2}{*}{$ < 28\times 10^{-11}$ @$90\%$ CL} & \multirow{2}{*}{\cite{KTeV:2003sls}} & \multirow{2}{*}{ 25\% \cite{JParcKaon2024}}\\
%%%
BR$(K_L\to \pi^0 e\bar e)(-)$         & $(1.55^{+0.60}_{-0.48})\times 10^{-11}$        &&& \\
%%%
BR$(K_L\to \pi^0 \mu\bar \mu)(+)$         & $(1.38^{+0.27}_{-0.25})\times 10^{-11}$    & \multirow{2}{*}{$ < 38\times 10^{-11}$ @$90\%$ CL} & \multirow{2}{*}{\cite{KTEV:2000ngj}} & \multirow{2}{*}{ 25\% \cite{JParcKaon2024}} \\
%%%
BR$(K_L\to \pi^0 \mu\bar \mu)(-)$         & $(0.94^{+0.21}_{-0.20})\times 10^{-11}$       &&&  \\
\hline \hline
\end{tabular}}
\caption{\small
The SM predictions, current experimental values, and projected precisions. In the last column, ``Current'' signifies that the measurement precision or the upper bound is maintained at the current experimental level.
\label{tab:data}}
\end{center}
\end{table}

\begin{figure}[b!]
\begin{center}
\includegraphics[width=0.48\textwidth]{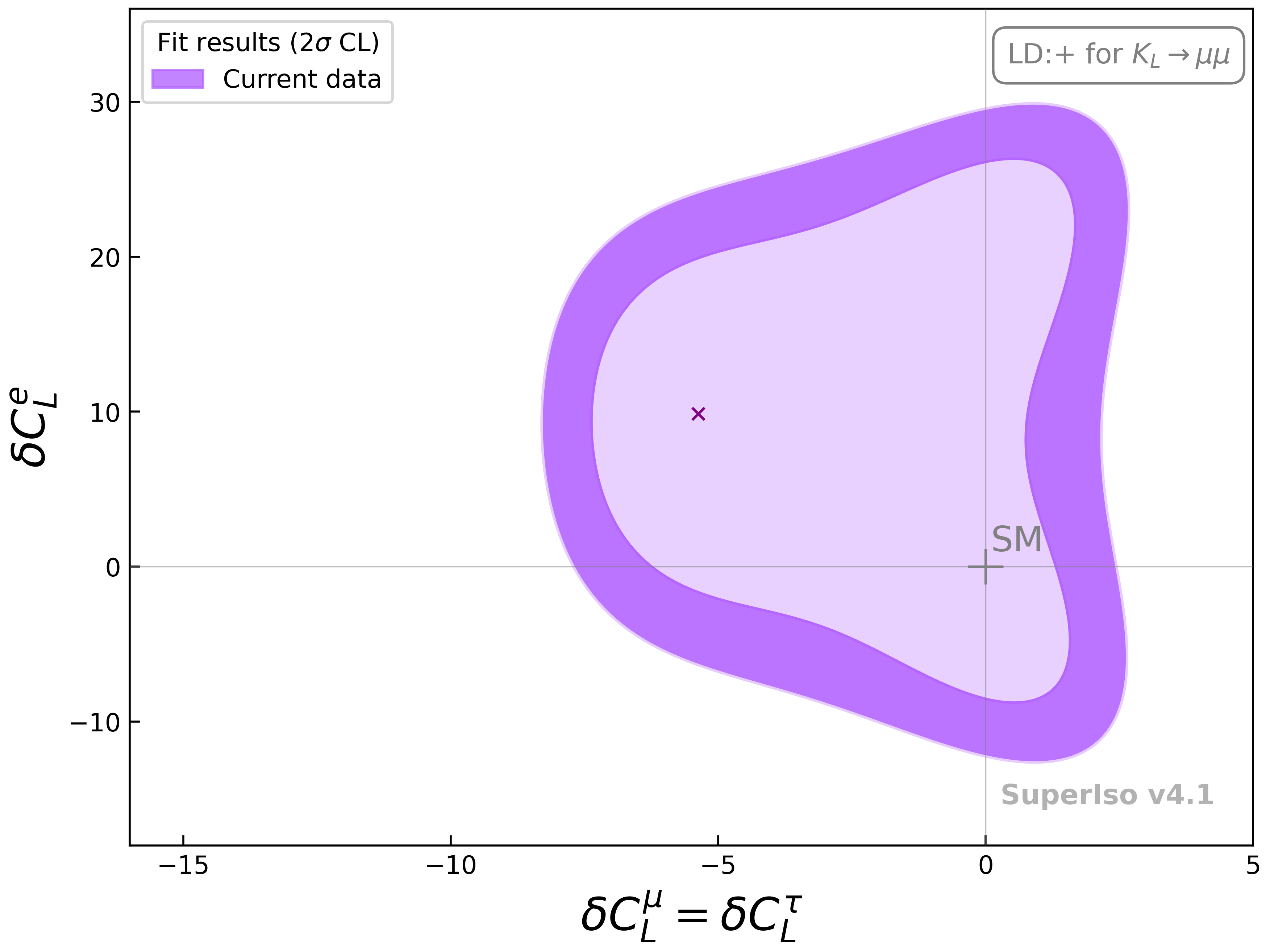}
\caption{\small
Results of the global fit for the scenario  $\delta C_L^{\tau}=\delta C_L^{\mu}$ with the best-fit point indicated by the purple cross.  The fit is implemented using the existing data with a positive signed long-distance contribution to $K_L\to \mu \bar{\mu}$.
\label{fig:fitpresent}}
\end{center}
\end{figure}

The impact of the current status of measurements on the Wilson coefficients is illustrated in Fig. \ref{fig:fitpresent}. For $K_L\rightarrow\mu\bar\mu$ we assume a positive sign for the long distance (LD) contributions. The light (dark) purple bands represent the $68\% \, (95\%)$ confidence interval regions in the $\delta C^e_L-\delta C^\mu_L$ plane and the purple cross represents the best-fit points.
\newpage
The numerical analysis is rendered more lucid by distributing it over three scenarios. In going from the first to the third, we sequentially add specific observables to the fit to illustrate their potential impacts.  \\
\\
 \textit{Scenario 1:} Projected 15\% precision for BR($K^+\rightarrow \pi^+\nu\bar\nu$) and 25\% precision for \mbox{BR($K_L\rightarrow \pi^0\nu\bar\nu$)}. \\
 \textit{Scenario 2:} Scenario 1 $+$ projected 25\% precision for BR($K_L\rightarrow \pi^0 e\bar e$). \\
 \textit{Scenario 3:} Scenario 2 $+$ projected 25\% precision for BR($K_L\rightarrow \pi^0 \mu\bar\mu$). \\

In the case of projective analyses, the choice of final measured central values for the observables mandates the adoption of reasonable assumptions. Since precise future measurements are unknown, we consider two illustrative scenarios to capture the range of possible outcomes. We invoke two possibilities as follows:
 \textbf{Projection A}: Observables already measured are kept, others assumed at their SM values, all with target precision of KOTO\mbox{-}II. \\
 \textbf{Projection B}: All measurements give current best-fit point with target precision of KOTO\mbox{-}II. The best-fit points are obtained from the analyses in Fig. \ref{fig:fitpresent}

\vspace*{0.3cm}
\section{Results and Discussion}
\label{sec:sec4}
With the theoretical and numerical methodology in place, we now present the inferences of the analysis for each of the three scenarios. 
The scenarios considered are based on estimated improved experimental precisions. However, we note that future measurements could achieve even higher precisions than currently anticipated; the impact of such improvements on the golden channels is illustrated in Fig.~\ref{fig:GoldenChannels_reducedErrs} in the appendix.
The findings reported in this paper are produced with the publicly available SuperIso program~\cite{Mahmoudi:2007vz,Mahmoudi:2008tp,Mahmoudi:2009zz,Neshatpour:2021nbn,Neshatpour:2022fak}.

\vspace*{0.2cm}
\subsection*{Scenario 1}

\begin{figure}[htb!]
\begin{center}
\includegraphics[width=0.48\textwidth]{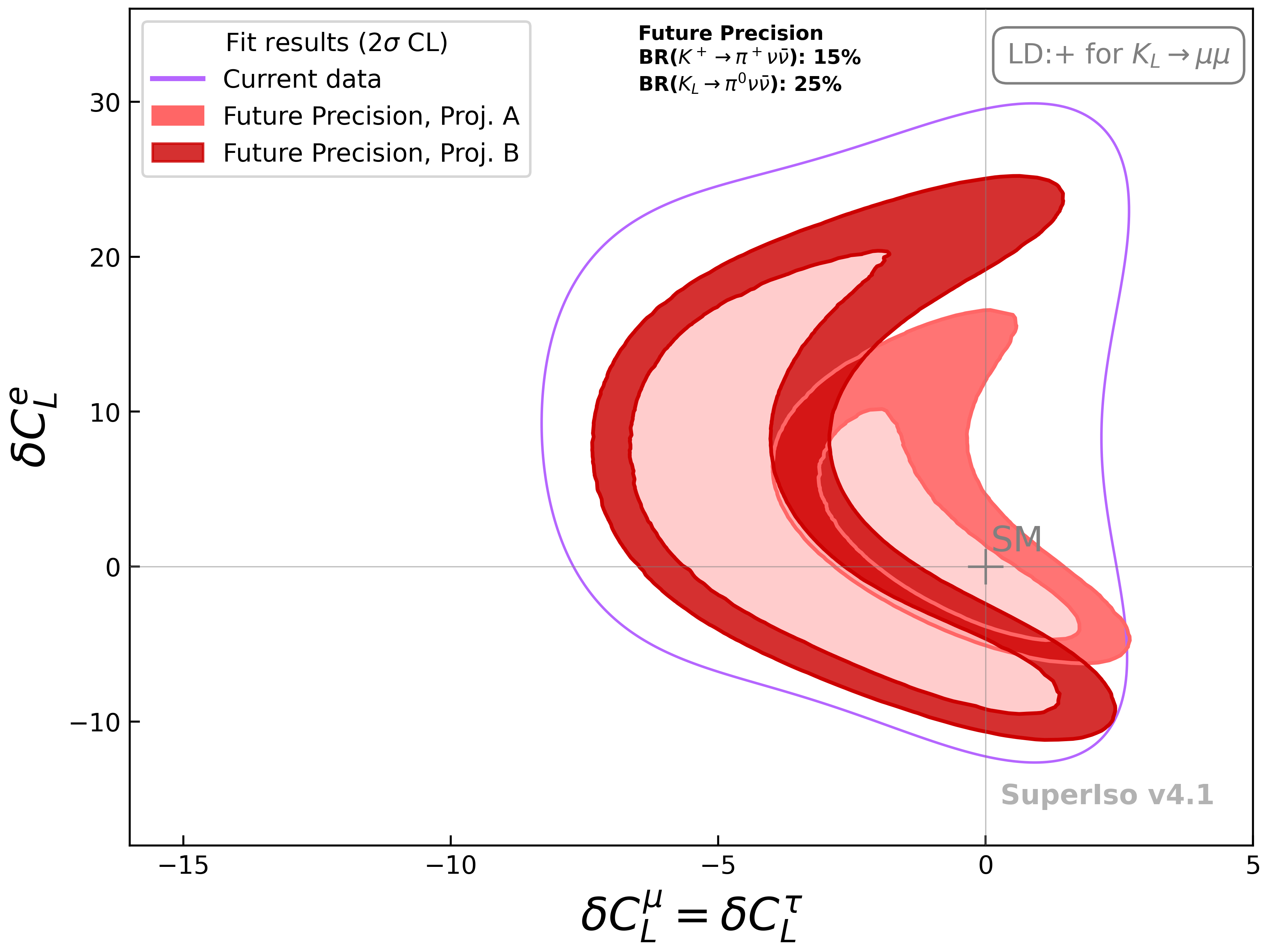}\\[0.2cm]
\includegraphics[width=0.48\textwidth]{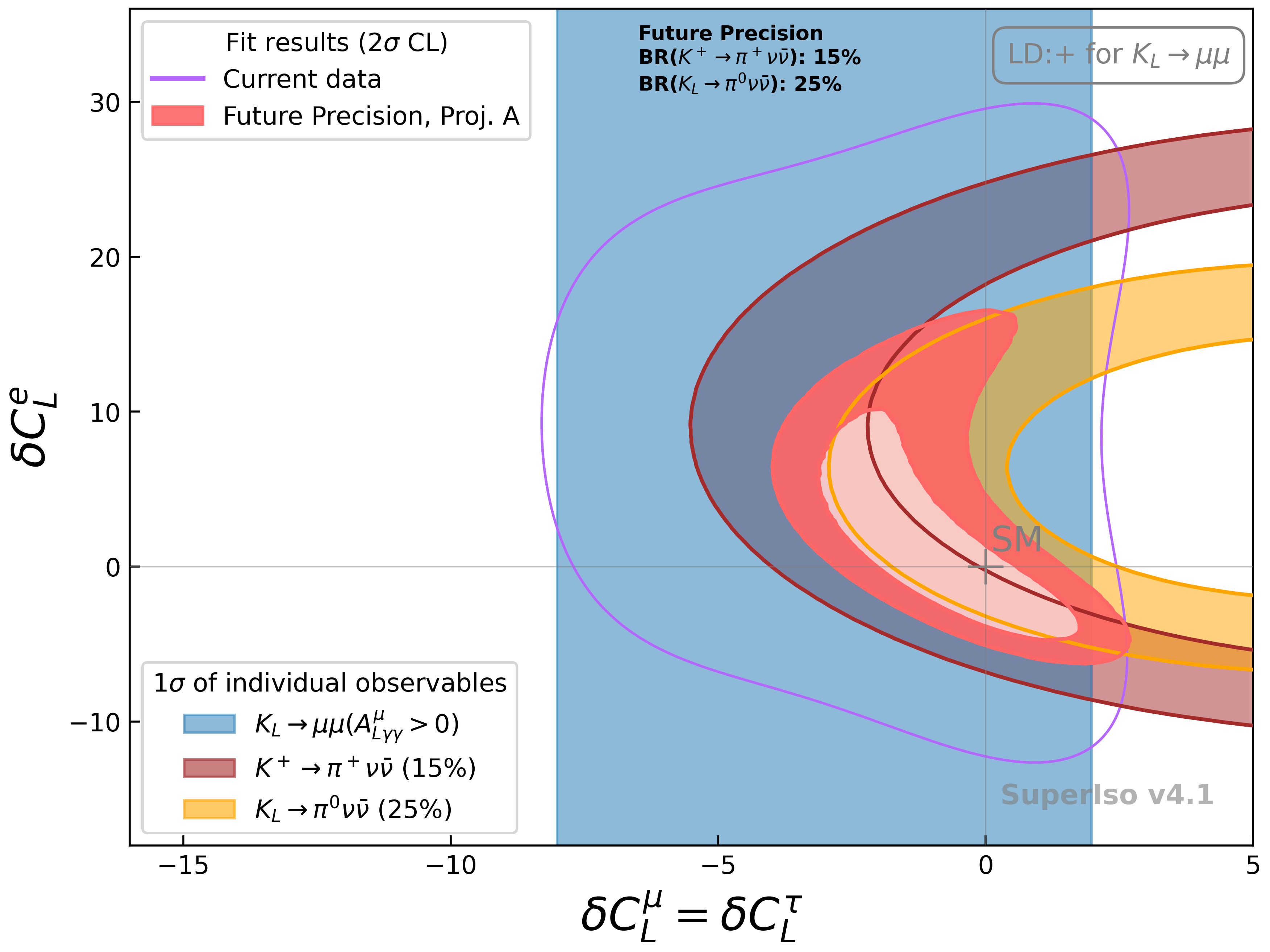}\quad
\includegraphics[width=0.48\textwidth]{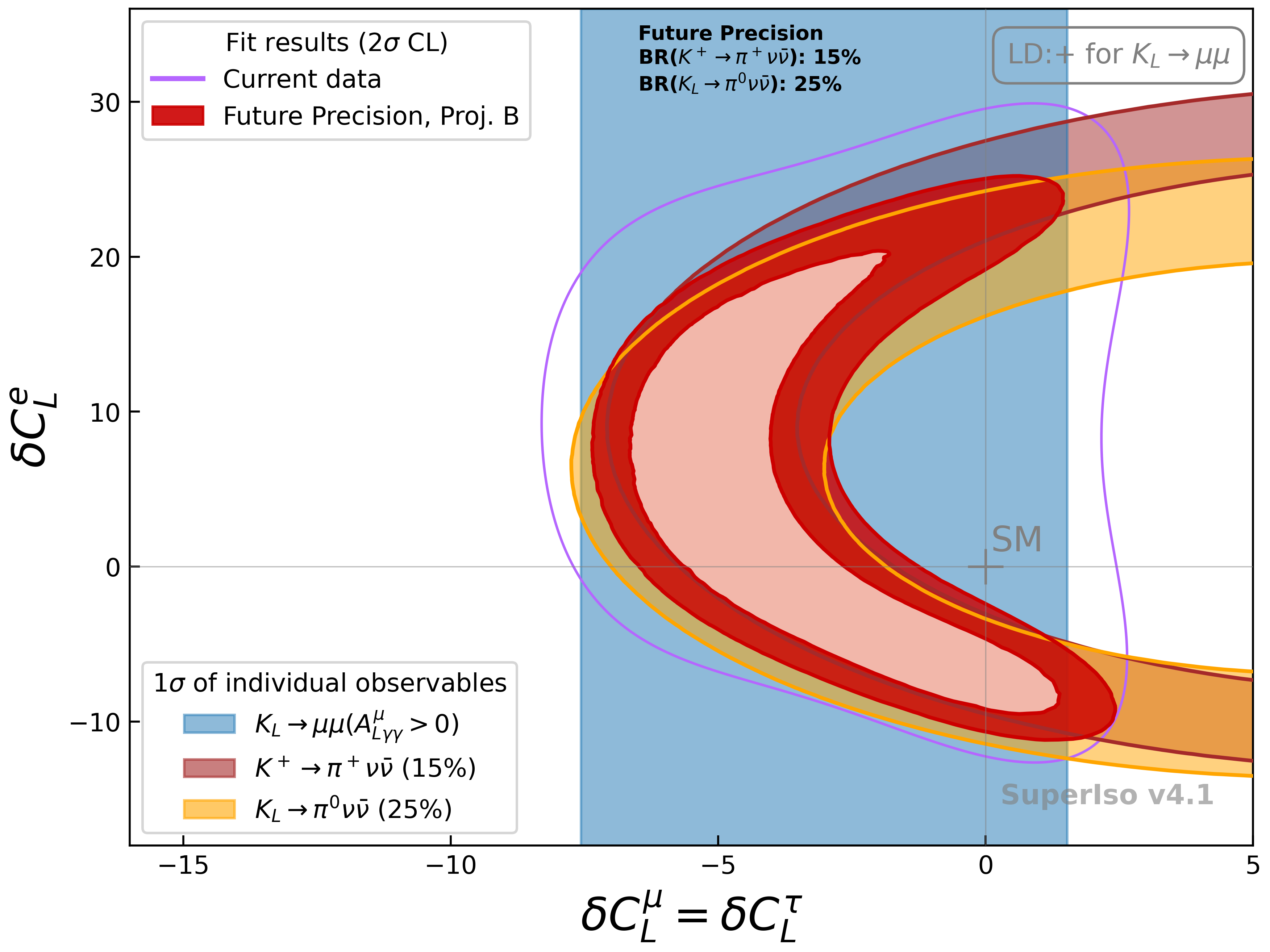}
\caption{\small Results corresponding to \textit{Scenario 1}. The top row illustrates the impact on parameter space with the consideration of the golden channels of kaon decays at their projected precisions. The dark (light) red represents $2\sigma$ CL regions for Projection A (B). The left (right) plot of the lower row gives the impact of the individual observables on the fit for Projection A (B).
\label{fig:scenario1}}
\end{center}
\end{figure}

The two golden channels in kaon physics are $K^+\rightarrow\pi^+\nu\bar\nu$ and $K_L\rightarrow\pi^0\nu\bar\nu$ owing to their sensitivity to short-distance physics. As a result, it would be instructive to see the extent of their impact on the $(\delta C^\mu_L,\delta C^e_L)$ parameter space at the end of their respective runs for data accumulation. 
For the case of $K^+\rightarrow\pi^+\nu\bar\nu$, we assume its measurement with $15\%$ precision at NA62 while $K_L\rightarrow\pi^0\nu\bar\nu$ is expected to be measured with $25\%$ precision at KOTO\mbox{-}II.
The top row of Fig. \ref{fig:scenario1} illustrates the extent of its impact in comparison to the existing status given by the pink contour. The two regions inside this contour correspond to the two projections under consideration. While there is an obvious reduction in the available parameter space, it is also instructive to see the extent to which the individual observables influence this reduction. The bottom row gives the impact of the potential measurements on the fits for Projection A (left) and Projection B (right). In the left plot, the allowed regions for $K_L\rightarrow\pi^0\nu\bar\nu$ (orange) and $K^+\rightarrow\pi^+\nu\bar\nu$ (maroon) are different as the former assumes a SM central value while the central value of the latter is the same as the current measurement at NA62 ($(10.6^{+4.0}_{-3.5} \pm 0.9 ) \times 10^{-11}$) which is about $1\sigma$ away from the SM prediction.  Since Projection B corresponds to using the best-fit values obtained from the existing fits as the measured central value for the observables, the orange and the maroon regions exhibit a larger overlap. In either  plot, inferring from the extent of overlap with the $2\sigma$ regions of the fit with the orange contour, suggests the importance of a precise $K_L\rightarrow\pi^0\nu\bar\nu$ measurement. The light blue regions are the regions allowed by BR$(K_L\rightarrow\mu\bar\mu)$.
The complete immersion of the $2\sigma$ regions inside this region indicates the weak influence of the decay modes on the final fits.

\newpage
\subsection*{Scenario 2}

\begin{figure}[htb!]
\begin{center}
\includegraphics[width=0.48\textwidth]{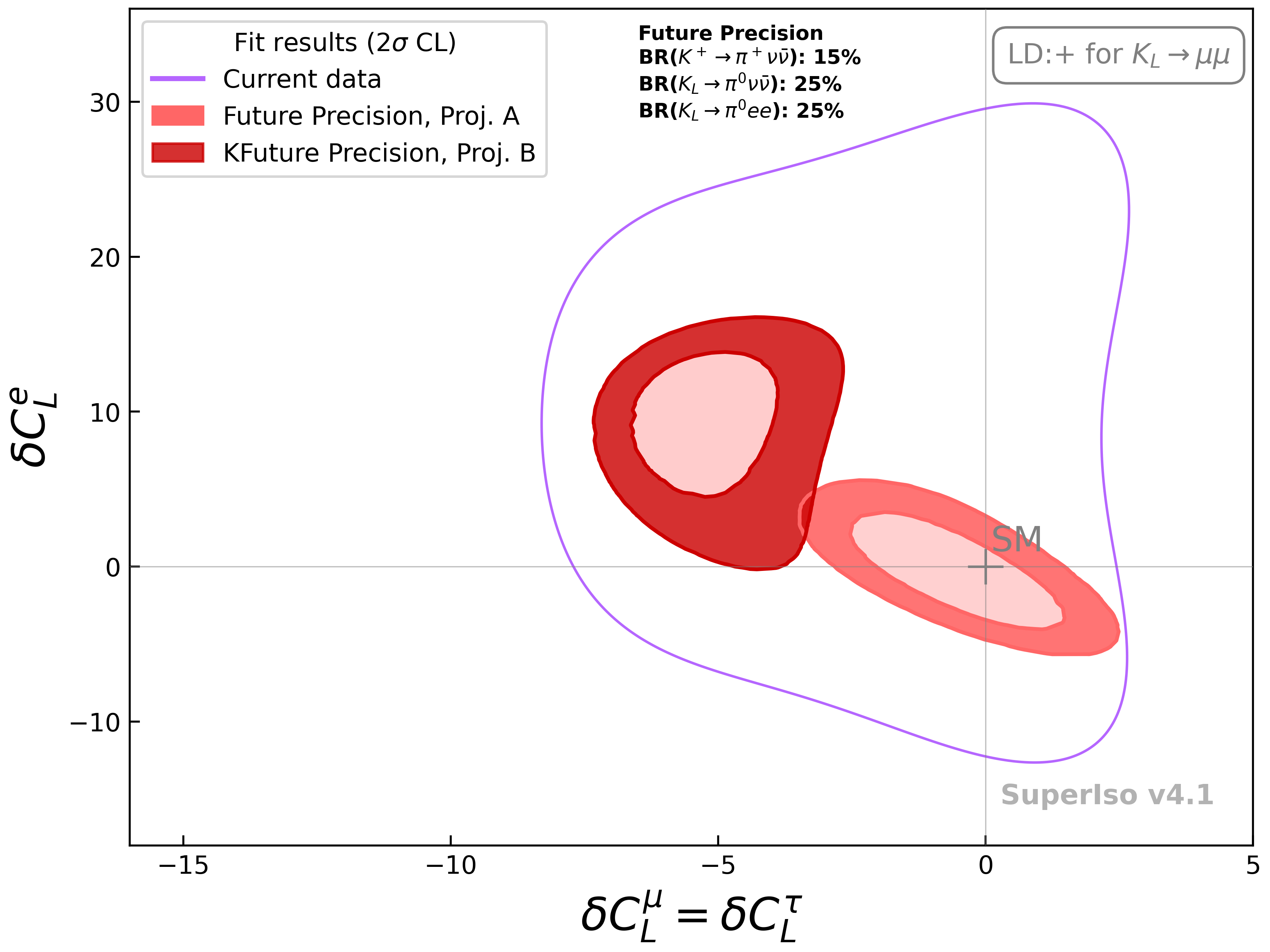}
\\[0.2cm]
\includegraphics[width=0.48\textwidth]{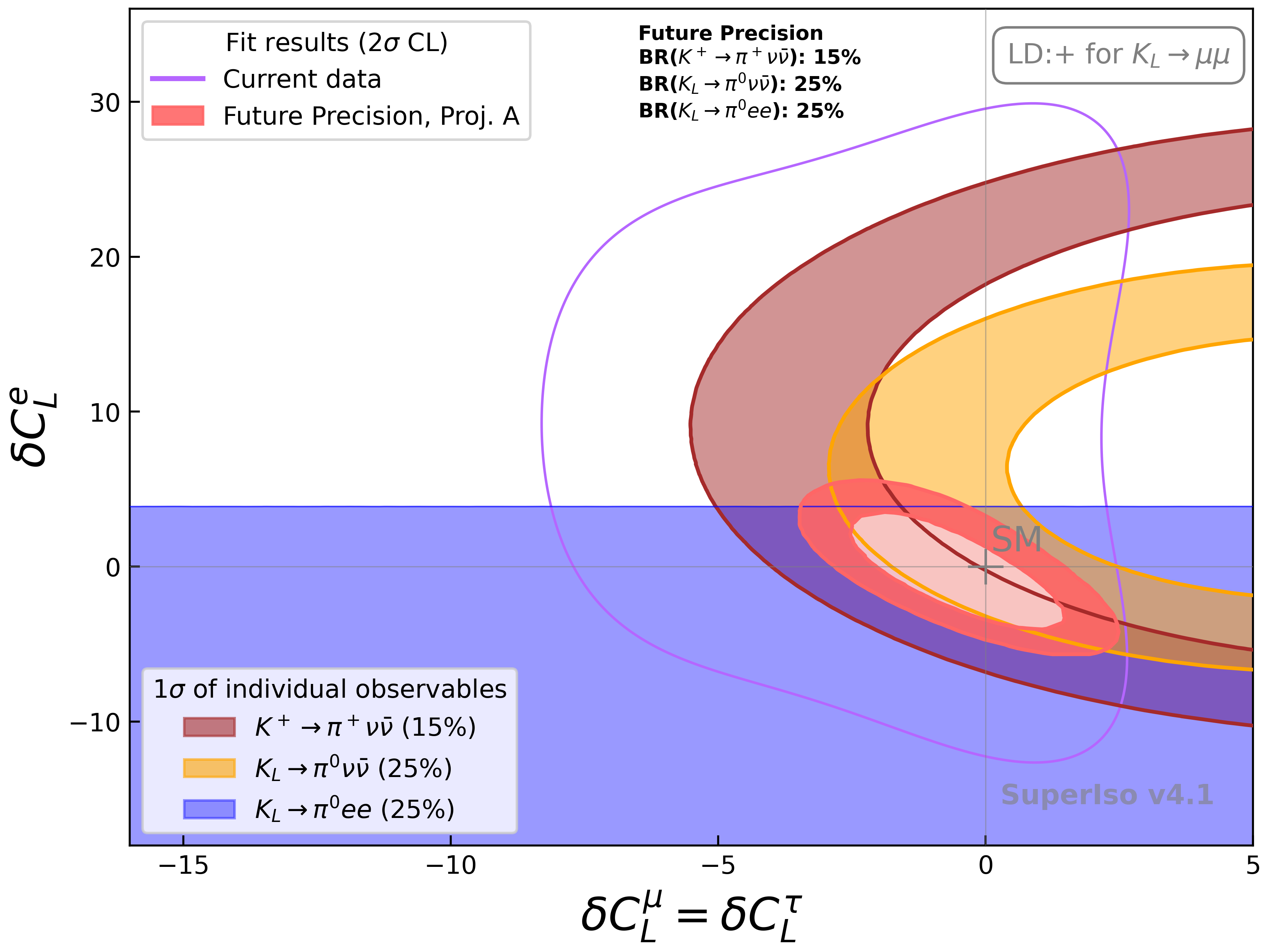}\quad\includegraphics[width=0.48\textwidth]{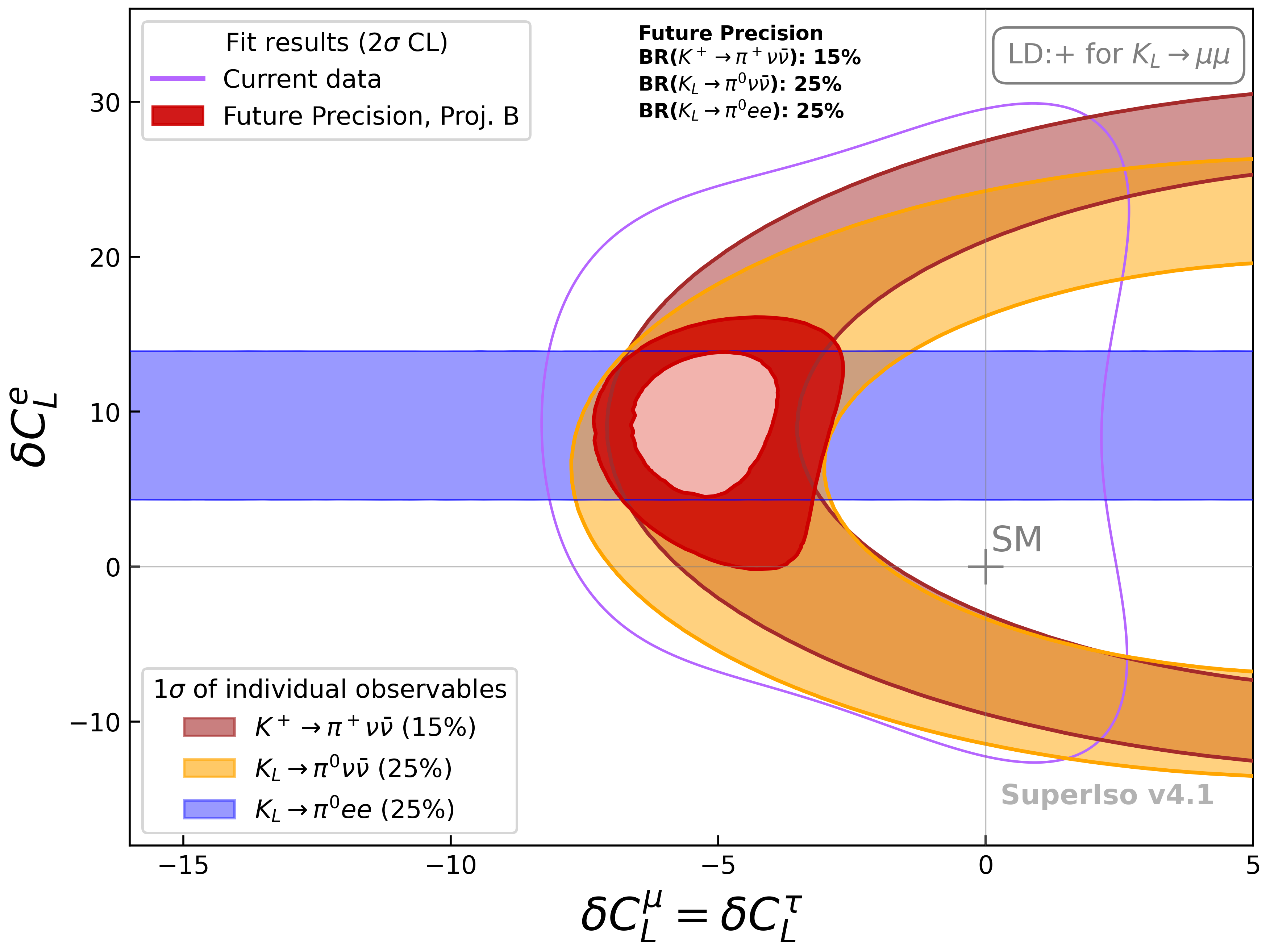}

\caption{\small Results corresponding to \textit{Scenario 2}. The top row illustrates the impact on the parameter space with also the inclusion of BR($K_L\rightarrow\pi^0 e\bar e$) to the fits. The dark (light) red represents $2\sigma$ CL regions for Projection A (B). The left (right) plot of the lower row gives the impact of the individual observables on the fit for Projection A (B).
\label{fig:scenario2}}
\end{center}
\end{figure}

This scenario is characterised by all the ingredients of Scenario 1 with the addition of a projected precision for $K_L\rightarrow\pi^0 e\bar e$. It is assumed that the branching fraction for this decay mode would be measured with $25\%$ precision, which is the same as that considered for $K_L\rightarrow\pi^0 \nu\bar \nu$. The $2\sigma$ regions corresponding to Projection A and B are given in the upper plot of Fig. \ref{fig:scenario2}. It illustrates a further reduction in the available parameter space in comparison to the corresponding plot in Fig. \ref{fig:scenario1}. The fact that this reduction is due to the addition of $K_L\rightarrow\pi^0 e\bar e$ is visible in the lower plots of Fig. \ref{fig:scenario2}. The dark blue regions represent the $1\sigma$ allowed regions of BR($K_L\rightarrow\pi^0 e\bar e$) for the corresponding projection. The fact that the available parameter space has moved so as to sit inside the dark blue region is suggestive of the impact of the measurement of BR($K_L\rightarrow\pi^0 e\bar e$) on the global fit.

\subsection*{Scenario 3}
\begin{figure}[htb!]
\begin{center}
\includegraphics[width=0.48\textwidth]{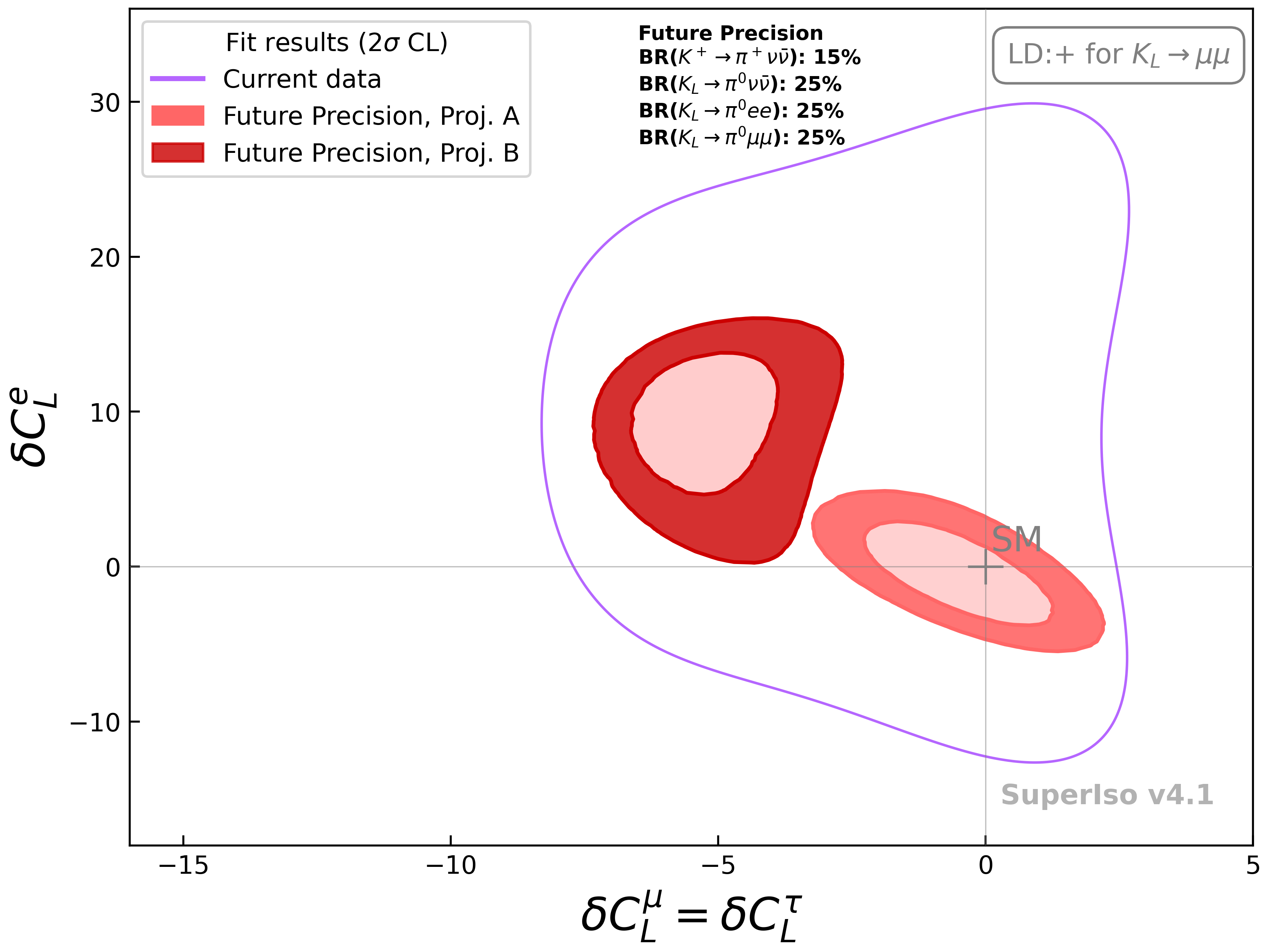}\\[0.1cm]
\includegraphics[width=0.48\textwidth]{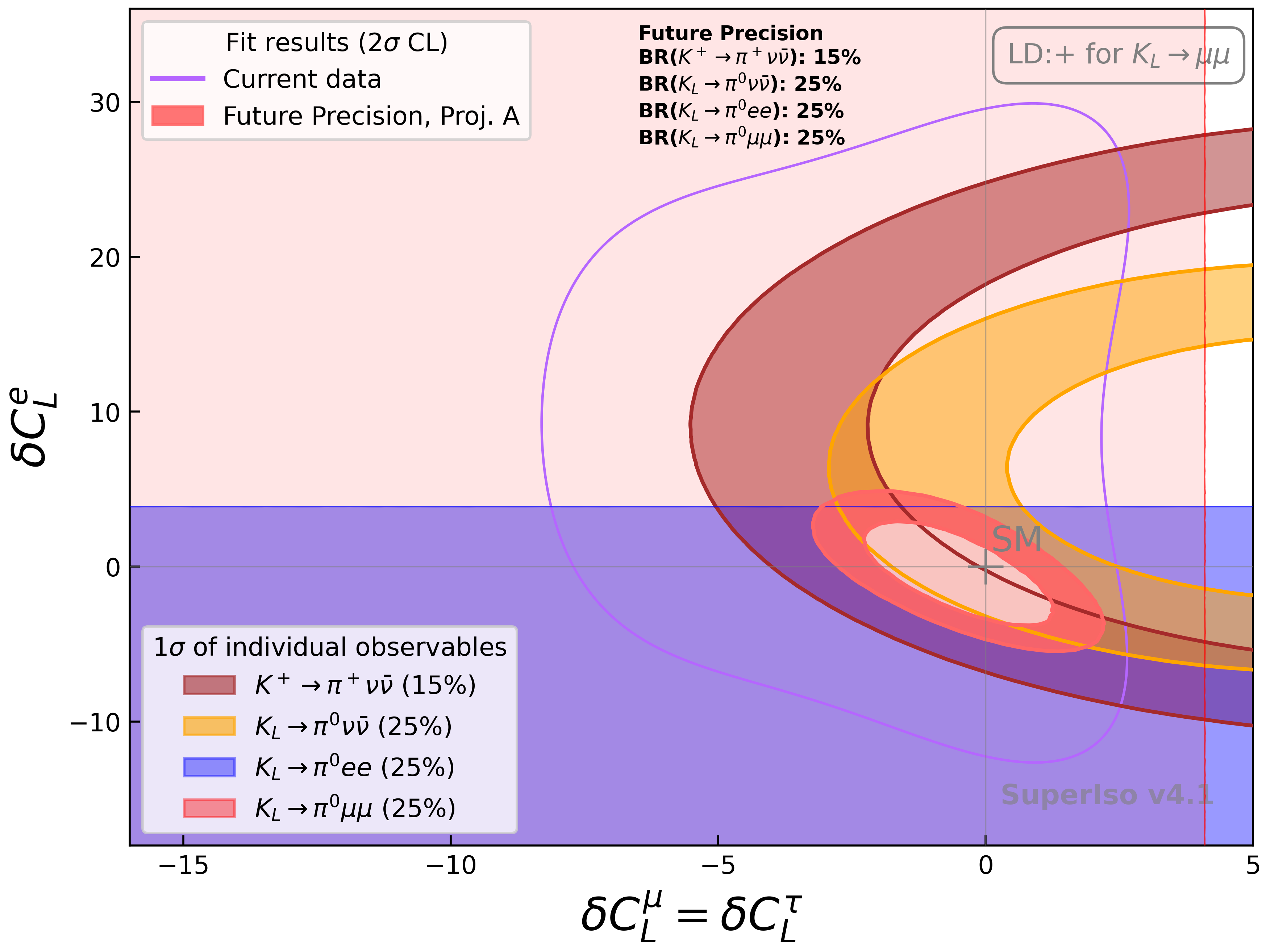}\quad \includegraphics[width=0.48\textwidth]{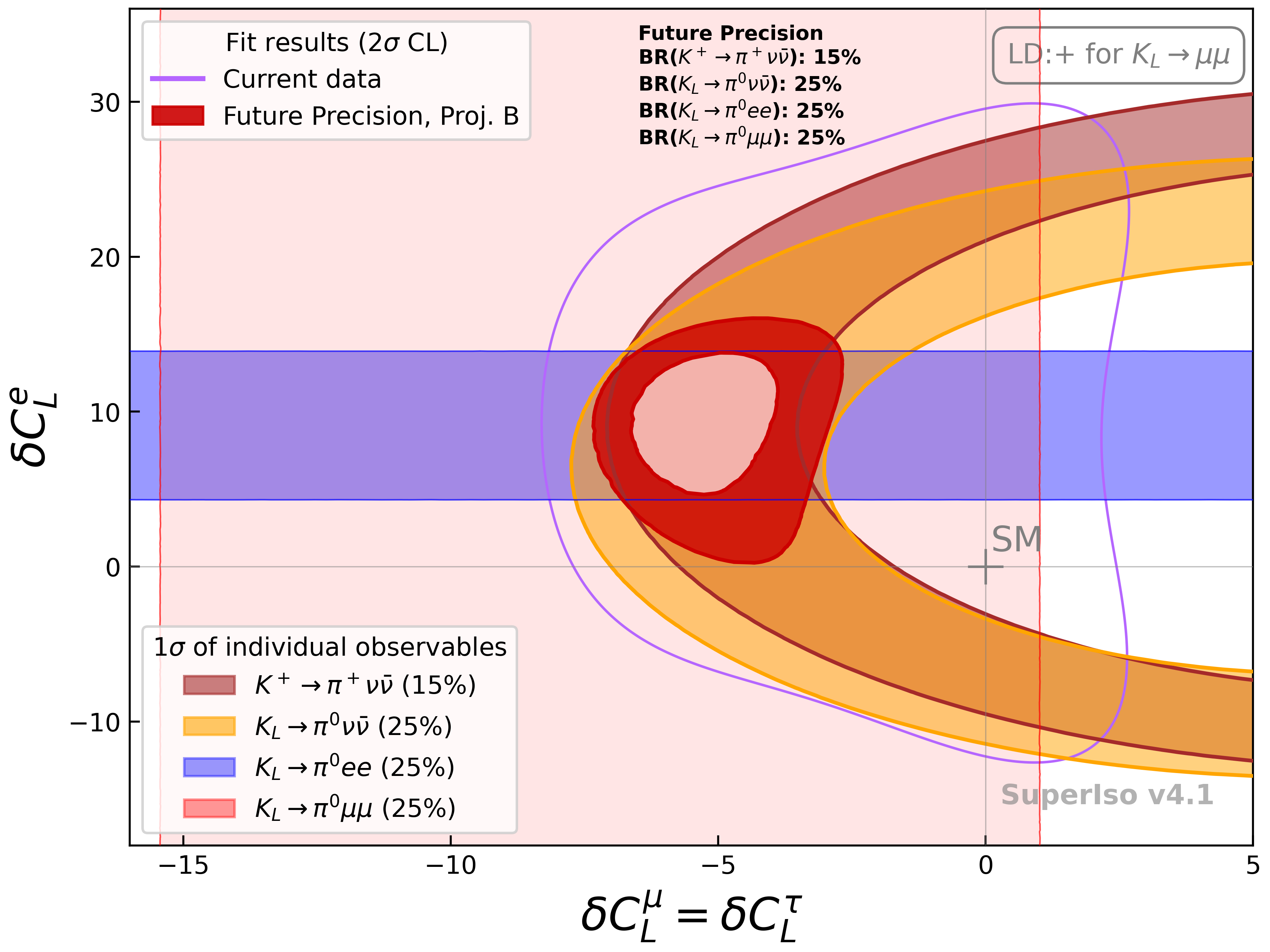}
\caption{\small Results corresponding to \textit{Scenario 3}. The top row illustrates the impact on parameters space with also the inclusion of BR($K_L\rightarrow\pi^0 \mu\bar\mu$) to the fits. The dark (light) red represents $2\sigma$ CL regions for Projection A (B). The left (right) plot of the lower row gives the impact of the individual observables on the fit for Projection A (B).
\label{fig:scenario3}}
\end{center}
\end{figure}

The final scenario builds on Scenario 2 by adding  the observable BR($K_L\rightarrow\pi^0 \mu\bar \mu$) to the fits. Similar to the other cases, it goes with the assumption that it will also be measured with $25\%$ precision. On studying the results of the fit in the top row of Fig.\ref{fig:scenario3}, it seems that there is no visual reduction in the available parameter space compared to Scenario 2. However, we find that this addition has the potential to separate the two Projections that are based on two different hypotheses: loosely they are SM-like and non-SM like. The overall result is quite significant as it also illustrates the power of precise measurements in kaon decays on the eventual conclusions. The bottom row of Fig. \ref{fig:scenario3}, gives the impact of the individual observables on the available regions for either projection. As the available region due to BR($K_L\rightarrow\pi^0 \mu\bar\mu$) is rather broad due to huge uncertainties, the dominating observable in this case is also BR($K_L\rightarrow\pi^0 e\bar e$) as seen from the overlap of the best-fit regions with the blue regions.

\section{Conclusions}
\label{sec:sec5}
The rare kaon decay $K_L\rightarrow\pi^0\ell\bar\ell$ is interesting as the current experimental bound is almost at the door of its SM value. In addition to being a testing ground for direct CP violation, the mode also has the possibility to have an influence on the acceptable parameter space of the Wilson coefficients. 
The paper focuses on tapping the potential of its measurement, with reasonable precision, by undertaking a systematic global analysis. The work is presented by means of dividing the numerical methodology into three scenarios, each of which is attributed to a certain choice of kaon decay observables. Going from \textit{Scenario 1-3}, it begins with the golden kaon channels and then the sequential addition of BR$(K_L\rightarrow\pi^0 e\bar e)$ and BR$(K_L\rightarrow\pi^0 \mu\bar \mu)$. The impact is already visible in \textit{Scenario 2}, which significantly narrows down the parameter space from existing status and \textit{Scenario 1} \textit{viz.} that uses the golden decay channels at their projected precisions. \textit{Scenario 2} with the addition of BR$(K_L\rightarrow\pi^0 \mu\bar \mu)$ facilitates the first clear distinction between a SM-like and a NP-like hypothesis. However, it also presents a need for a detailed theoretical effort into the calculation of the SM values and reduction of the corresponding errors. In general, the study presents a strong case for pursuing a kaon program with a specific focus on $K_L$ decays. With the ongoing measurement of BR($K_L\rightarrow\pi^0 \nu\bar\nu$) at KOTO, its upgrade to KOTO\mbox{-}II offers an excellent opportunity to measure the decay mode into both the neutrinos and the charged leptons. Thus, a new dawn for kaon physics is essential to view the Standard Model with a new light.

\section*{Acknowledgements}
We appreciate the valuable input from E.~Goudzovski, M.~Koval, and C. Lazzeroni of the NA62 collaboration and Diego Martinez Santos from the LHCb collaboration. We thank Avital Dery for discussions on $K^0\rightarrow\mu\bar\mu$.
This research was funded in part by the National Research Agency (ANR) under project no. ANR-21-CE31-0002-01. AMI would like to acknowledge the generous support by SERB India through project no. SRG/2022/001003. GD was supported in part by the INFN research initiative Exploring New Physics (ENP).

\begin{appendix}
\section{Further Discussions for the Golden Channels}
\begin{figure}[t!]
\begin{center}
\includegraphics[width=0.48\textwidth]{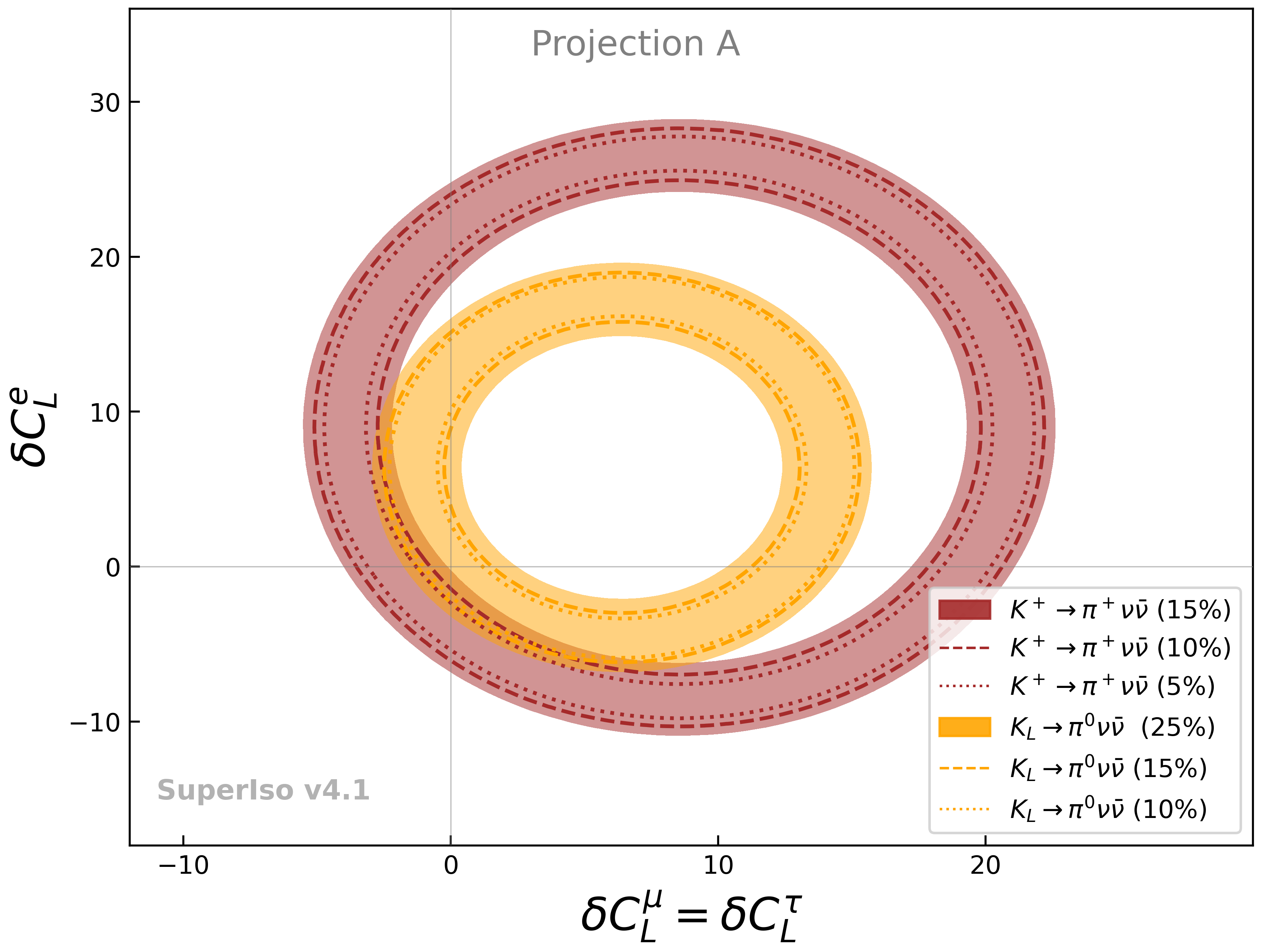}\quad \includegraphics[width=0.48\textwidth]{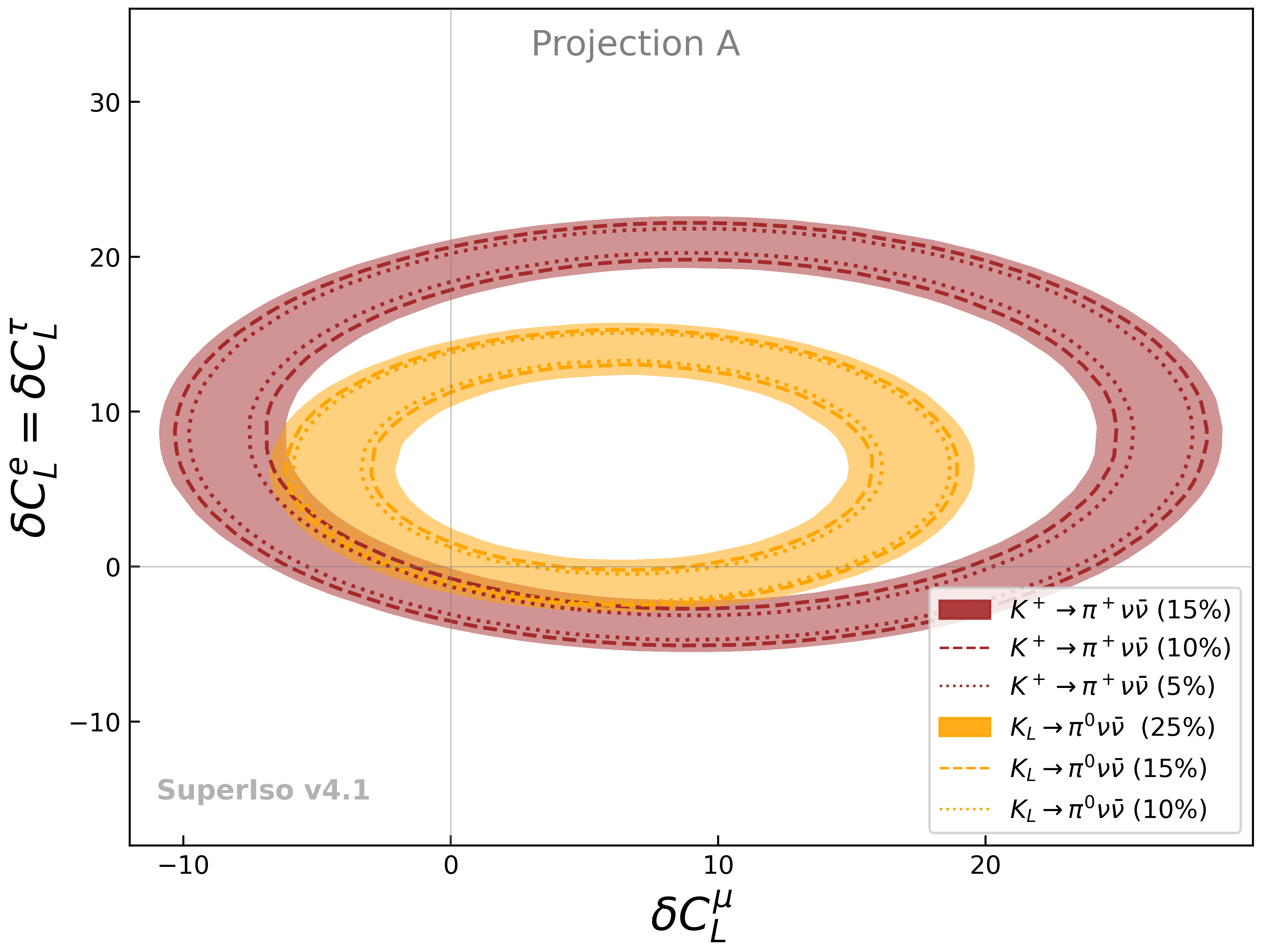}
\caption{\small 
Results corresponding to \textbf{Projection A}. In the left plot, New Physics contributions respect lepton flavour universality (LFU) in the muon and tau channels, while in the right plot, LFU is preserved between the electron and tau channels. The colored regions represent projected precisions of 15\% and 25\% for the \( K^+ \to \pi^+ \nu \bar{\nu} \) and \( K_L \to \pi^0 \nu \bar{\nu} \) decays, respectively, with dashed and dotted contours indicating improved projections of 10\% and 5\% for the former and 15\% and 10\% for the latter.
\label{fig:GoldenChannels_reducedErrs}}
\end{center}
\end{figure}

To illustrate the impact of reduced uncertainties in the golden channels, Fig.~\ref{fig:GoldenChannels_reducedErrs} shows the $1\sigma$ CL for \textbf{Projection A} assuming uncertainties of 15\%, 10\%, and 5\% for \( K^+ \to \pi^+ \nu \bar{\nu} \) and 25\%, 15\%, and 10\% for \( K_L \to \pi^0 \nu \bar{\nu} \). The left plot follows the same assumptions as the main text regarding LFUV contributions, where \( \delta C_L^e \neq \delta C_L^\mu = \delta C_L^\tau \), while the right plot shows results for \( \delta C_L^\mu \neq \delta C_L^e = \delta C_L^\tau \).

It is worth noting that these projections assume no improvements in theoretical precisions. This explains why, for instance, reducing the experimental uncertainty of $\text{BR}(K^+ \to \pi^+ \nu\bar\nu)$ from 15\% to 5\% has a somewhat modest effect on shrinking the $1\sigma$ CL region, given the current theory uncertainty of around 8\%. This is also seen for $\text{BR}(K_L \to \pi^0 \nu\bar\nu)$, where the theoretical uncertainty is approximately 11\%. Nonetheless, future theoretical advancements are expected, as the primary source of uncertainty stems from CKM parameters.

\end{appendix}

\bibliographystyle{JHEP} 
\bibliography{biblio}

\clearpage

\end{document}